\documentstyle[preprint,prl,aps,epsf]{revtex}

\title{Phase space dynamics of overdamped quantum systems} 

\author{Joachim Ankerhold}
\address{Physikalisches Institut, Albert-Ludwigs-Universit{\"a}t
Freiburg,
 Hermann-Herder-Stra{\ss}e 3, D-79104 Freiburg, Germany}

\date{\today}
\begin{document}
\maketitle

\begin{abstract}
The phase space dynamics of dissipative quantum systems in strongly condensed
phase is considered. Based on the exact path integral approach it is
shown that the Wigner transform of the reduced density matrix obeys a
time evolution equation of Fokker-Planck type valid from high down to
very low temperatures. The effect of quantum fluctuations is
discussed and the accuracy of these findings is tested against exact data
for a harmonic system.

\end{abstract}

{\it Introduction--}Dynamics in strongly condensed phase can be found almost
everywhere in nature, e.g.\ for tunnel diodes in mesoscopic physics,
for macromolecules  in biological and soft matter systems, and for chemical
reactions, to name but a few. Associated with this bunch of
realizations is a rich phenomenology comprising prominent effects such
as stochastic resonance \cite{jung}, resonant activation \cite{fluc1},
transport in ratchets \cite{reimann}, and
adiabatic electron 
transfer \cite{garg} which have been explored extensively in the last decade.
However, most of these studies have focused on the domain of
classical physics and less is known about low temperature quantum
properties. This is mainly due to the fact that in contrast to the
Fokker-Planck
equation for the classical phase space distribution \cite{risken}, a
simple time 
evolution equation 
for the reduced density matrix of a dissipative quantum system does
not exist in general \cite{weiss}. Typically, quantum fluctuations appear on a
time scale $\hbar\beta$ ($\beta=1/k_{\rm B} T$) so that at lower
temperatures the quantum stochastic process becomes strongly
non-Markovian and intimately depends on the initial correlations
between system and heat bath. An exact expression for the density
matrix is available within the path integral formulation, but
numerical evaluations are in most cases prohibitive \cite{weiss}.

In the low friction limit, e.g.\ for quantum optical systems, progress
can be made by a Born-Markov approximation leading to well-known
master equations. The opposite limit of strong damping has only
recently attracted 
considerable attention
\cite{thorwart}--\cite{baner}. In particular, it was
shown \cite{ankerhold} that for strong 
friction and low temperatures the density matrix is essentially
restricted to its diagonal part, the position probability
distribution, and that this part obeys a quantum analogue of the
classical Smoluchowski equation. The question we
address here goes far 
beyond: Is there a time evolution equation in whole phase space valid
from high down to low temperatures provided friction is sufficiently
strong? The answer is yes and we will show this in detail below. 

{\it Path integral formulation--}The inclusion of dissipation
within quantum mechanics is well established and is based on  a
system+reservoir 
formulation \cite{weiss}. The dynamics of the corresponding density
matrix starting 
at $t=0$ from a general initial state $W(0)$
 reads
\begin{equation}
W(t)= \exp(-i H t/\hbar)\, W(0)\, \exp(i H t/\hbar)\label{eq1}
\end{equation}
where the Hamiltonian $H=H_S+H_R+H_I$ contains a system, a
reservoir (heat bath), and a system-reservoir interaction
part, respectively. Dissipation arises as the effective impact of the
reservoir degrees of freedom on the system dynamics within a reduced
picture $\rho(t)=tr_R\{W(t)\}$. In the standard description 
the Gaussian statistics of the heat bath is modeled by a quasi-continuum of
harmonic oscillators bilinearly coupled with the system. In fact, one
this way regains in the classical limit the generalized Langevin
equation of Brownian motion. The only non-perturbative treatment
of the system-reservoir coupling is an exact elimination of the
bath degrees of freedom by means of the path integral approach. For
the position representation of the reduced density matrix one obtains
\begin{equation}
\rho(q_f,q_f',t)=\int dq_i\, dq_i'\ J(q_f,q_f',t,q_i,q_i')\
\lambda(q_i,q_i'). \label{eq2}
\end{equation}
Here, the propagating function $J(\cdot)$ is a threefold path integral over
the system degrees of freedom only. The two real time paths $q(s)$ 
and $q'(s)$ connect in time $t$ the 
initial points $q_i$ and $q_i'$ with the fixed end points $q_f$ and
$q_f'$, while the imaginary time path $\bar{q}$ runs from $q_i$ to
$q_i'$ in the interval $\hbar\beta$.  
The contribution of each path is weighted by
$\exp(i\Sigma[q,q',\bar{q}]/\hbar)$ with an effective action
$\Sigma[q,q',\bar{q}]$ not specified here explicitly.
Basically, it comprises the actions of the bare system in imaginary
and real time, respectively, and additional interaction contributions
(influence 
functional), non-local in 
time, which in the reduced picture rule the influence of the reservoir
onto the system. In the limit $t\to 0$ one has
$J(q_f,q_f',t,q_i,q_i')\to \rho_\beta(q_i,q_i')\, \delta(q_f-q_i)\,
\delta(q_f'-q_i')$ so that
\begin{equation}
\rho(q_f,q_f',0)=\rho_\beta(q_f,q_f')\ \lambda(q_f,q_f')\label{eq4}
\end{equation}
with the reduced equilibrium density matrix $\rho_\beta(q,q')=\langle
q|tr_R \exp(-\beta H)|q'\rangle$ and a preparation function $\lambda$
characterizing initial deviations from thermal equilibrium. The
crucial point here is that Eq.~(\ref{eq4}) is a correlated initial
state \cite{grabert}, in contrast to a factorizing initial state used
in ordinary 
Feynman-Vernon theory. In the strong friction regime the latter one is
{\em not} applicable in contrast to what has been done in
\cite{baner}. Even in the classical limit does a product 
initial state not allow for a consistent derivation of the well-known
Smoluchowski equation. 

{\it Overdamped quantum systems--}The stochastic dynamics of
dissipative quantum systems determined by Eqs.~(\ref{eq2}) and
(\ref{eq4}) is rather subtle as it strongly depends on
initial correlations between system and reservoir as well as on the non-local
self-interactions contained in the influence 
functional \cite{grabert}. Both are governed by the damping kernel
\begin{equation}
K(\theta)=\int_0^\infty \frac{d\omega}{\pi}\, I(\omega)\,
 \frac{\cosh[\omega(\hbar\beta/2-i\theta)]}{\sinh(\omega\hbar\beta/2)}
 \label{eqk}
\end{equation}
where $\theta=s-i\tau$, $0\leq s\leq t$, $0\leq \tau\leq \hbar\beta$
and $I(\omega)$ is the spectral density of the heat bath.
Accordingly, the impact of the bath is completely specified by
temperature $T$ and $I(\omega)$. In particular, the spectral density
is related to the  
macroscopic damping via
\begin{equation}
\gamma(s)=\frac{2}{M}\, \int_0^\infty \frac{d\omega}{\pi}\,
\frac{I(\omega)}{\omega}\, \cos(\omega s)\, .\label{eqd}
\end{equation} 

Now, we turn to the simplifications that
arise in the overdamped regime. For this purpose we define a typical
damping strength as
\begin{equation}
\gamma\equiv \hat{\gamma}(0)=\lim_{\omega\to 0}\,
\frac{I(\omega)}{M\omega}\label{eqgam}
\end{equation}
with $\hat{\gamma}(\omega)$ the Laplace transform of $\gamma(t)$. For
instance, in the 
ohmic case $I(\omega)=M \bar{\gamma} \omega$ and also
for the more realistic Drude model
$I(\omega)=M \bar{\gamma}\omega_c^2/(\omega^2+\omega_c^2)$ 
(cut-off frequency $\omega_c$)  one finds
$\gamma=\bar{\gamma}$.
Given a typical frequency $\omega_0$ of the bare
system, e.g.\ the ground state frequency, by
strong damping we then mean 
\begin{equation}
\frac{\gamma}{\omega_0^2}\gg \hbar\beta, \frac{1}{\omega_c},
\frac{1}{\gamma}.\label{eqcon}
\end{equation}
In other words, we assume the time scale separation well-known from the
classical overdamped regime \cite{risken} and extend it to the quantum
range by incorporating the time scale for quantum fluctuations
$\hbar\beta$. Correspondingly, we consider the dynamics
Eq.~(\ref{eq2}) on 
the coarse grained time scale
$s\gg \hbar\beta, \frac{1}{\omega_c},
\frac{1}{\gamma}$ and $\sigma\gg \frac{1}{\omega_c},
\frac{1}{\gamma}$.
The consequences are substantial: (i) the strong friction
suppresses non-diagonal elements of the
reduced density matrix during the time evolution. This simply reflects
the fact that a quantum system  
behaves more classically, the stronger coherences are destroyed by the
presence of a heat bath. (ii) The real-time part
$K(s)$ of the damping kernel becomes local on the coarse grained time
scale so that a time evolution equation of the form
$\dot{\rho}(t)={\cal L}\, \rho(t)$ with a time independent operator
${\cal L}$ may exist.

Following the above simplifications the path integral formulation
now allows for a perturbative treatment in the {\em strong damping
limit}. To this end it is convenient to
introduce sum and difference coordinates for the imaginary time path,
$\bar{r}=(\bar{q}+\bar{q}')/2$ and $\bar{x}=\bar{q}-\bar{q}'$, and sum
and difference paths in real-time, $r(s)=[q(s)+q'(s)]/2$ and
$x(s)=q(s)-q'(s)$, respectively. The idea is to evaluate the path
integrals in the sense of a semiclassical approximation by assuming
self-consistently that non-diagonal elements, i.e.\ $\bar{x}$ and
$x(s)$ dependent terms, remain small during the time evolution.
Hence  the effective action  
$\Sigma[r,x,\bar{q}]$ is expanded up to second order
in the $\bar{x}$ coordinate of the imaginary time path and in the
excursions $x(s)$ of the real-time path integrals. Doing so we take
sufficiently smooth potentials for granted.
While this procedure is applicable to a wide range of spectral bath
densities, we 
concentrate in the sequel on the quasi-ohmic case with a very large
cut-off frequency $\omega_c\gg \gamma$. It is worthwhile to
note that we do not restrict the value of $\gamma\hbar\beta$ meaning that our
analysis covers a broad  temperature range from the classical
($\gamma\hbar\beta\ll 1$) to the deep quantum 
domain ($\gamma\hbar\beta\gg 1$).

{\it Equilibrium density matrix--}The question to what extent the
thermodynamic equilibrium is affected 
by the strong damping limit, particularly at lower temperatures, is a
very crucial one and also serves as a basis for the more 
involved treatment of the dynamical case. In terms of a sum over paths
from $\bar{q}$ to $\bar{q}'$ in the time interval $[0,\hbar\beta]$
the unnormalized reduced equilibrium density matrix in position
representation reads  
\begin{equation}
\rho_\beta(\bar{q},\bar{q}')=\int {\cal D}[\bar{q}]\ {\rm
e}^{-S_E[\bar{q}]/\hbar-\phi_E[\bar{q}]/\hbar}\label{equi1}
\end{equation}
with the bare Euclidian action $S_E[\bar{q}]=\int_0^{\hbar\beta} d\tau\,
[M\dot{\bar{q}}^2/2+V(\bar{q})]$ of a particle of mass $M$ in the
potential $V(\bar{q})$ and the Euclidian influence functional
$\phi_E[\bar{q}]=\int_0^{\hbar\beta} d\tau\int_0^{\hbar\beta} d\sigma
K(i\tau-i\sigma)\, \bar{q}(\tau)\, \bar{q}(\sigma)$. 
To solve the path
integral we put
$\bar{q}=\bar{q}_{\rm ma}+\delta\bar{q}$ where $\bar{q}_{\rm
ma}(\sigma)$ is the minimal action path to (\ref{equi1})
and obeys the boundary conditions $\bar{q}(0)=\bar{q}$,
$\bar{q}(\hbar\beta)=\bar{q}'$, while $\delta\bar{q}(\sigma)$ denote quantum
fluctuations with $\delta\bar{q}(0)=\delta\bar{q}(\hbar\beta)=0$. The
most convenient way to calculate $\bar{q}_{\rm ma}$ is to switch to 
Fourier space with respect to the Matsubara frequencies $\nu_n=n
2\pi/\hbar\beta$. In the corresponding equation of motion to leading
order the friction
term prevails and one finds $\bar{q}_{\rm ma}=\bar{r}$ (recall that
$\bar{x}$ is 
assumed to be small, namely, of order $1/\sqrt{\gamma}$). Dynamical
contributions, however, must be taken into 
account if we are interested in next order corrections. The final
result for $\bar{q}_{\rm ma}$ including corrections of order $1/\gamma$
is lengthy and thus omitted here; the  minimal action
takes the form
\begin{equation}
S_E(\bar{r},\bar{x})/\hbar=\beta V(\bar{r})-\Lambda\beta^2\,
V'(\bar{r})^2+\frac{\Omega}{2\hbar^2}\,
\bar{x}^2+O(\Lambda/\gamma)
\label{minac}
\end{equation}
with
\begin{equation}
\Lambda=\frac{2}{M\beta}\,
\sum_{n=1}^\infty\frac{1}{\nu_n^2+\nu_n\hat{\gamma}(\nu_n)}\ \ \ \ ,\
\ \ \ \ 
\Omega=\frac{M}{\beta}+\frac{2
M}{\beta}
\sum_{n=1}^\infty\frac{\hat{\gamma}(\nu_n)}{\nu_n+\hat{\gamma}(\nu_n)}\,
.
\label{coeff}
\end{equation}
Apparently, $\Lambda$ measures the typical strength of quantum
fluctuations in position 
space, while $\Omega$ is via the identification $\bar{x}/\hbar\to p$ 
 associated with
the variance in momentum $\Omega=\langle p^2\rangle$. In case of Drude
damping with a high frequency cut-off $\omega_c$ both $\Lambda$ and
$\Omega$ can be expressed in terms of $\Psi$ functions. Then, for high
temperatures $\gamma\hbar\beta\ll 1$ we find $\Lambda\approx
\hbar^2\beta/12 M$ and $\Omega\approx M/\beta$.
The friction dependence appears as a genuine quantum effect for
lower temperatures and for $\gamma\hbar\beta\gg 1$ one has $\Lambda\approx 
(\hbar/M\gamma\pi) \log(\gamma\hbar\beta/2\pi)$ and $\Omega\approx
(M\hbar\gamma/\pi)\log(\omega_c/\gamma)$. With increasing
$\gamma$ the strong squeezing
of quantum fluctuations in
position induces enhanced fluctuations in
the momentum (see fig.~\ref{fig1}), thus suppressing nondiagonal
elements in the density matrix.  

By expanding the 
full action in (\ref{equi1}) up to second order in
$\delta\bar{q}$, the contribution of quantum fluctuations to
(\ref{equi1}) is obtained. Due to the strong friction higher order
terms are 
negligible. The corresponding Gaussian integral is again calculated by
utilizing a Fourier expansion in the Matsubara
frequencies. Eventually, the equilibrium density matrix in the strong
friction limit is found as
\begin{equation}
\rho_\beta(\bar{x},\bar{r})= \frac{1}{Z} \, {\rm
e}^{-\beta V(\bar{r})-\Omega\, \bar{x}^2/2\hbar^2}\ {\rm
e}^{\Lambda\beta [ \beta V'(\bar{r})^2/2-3 V''(\bar{r})/2]}\, \label{equiden}
\end{equation}
where $Z$ denotes a proper normalization factor, e.g.\ $Z=\int dq
\rho_\beta(0,q)$. 
Interestingly, the probability distribution
is Gaussian in $\bar{x}$, i.e. its Wigner transform ($\bar{x}/\hbar\to
p$) Gaussian in
momentum, even at low temperatures. Anharmonic corrections in
$\bar{x}$ to the exponent are at most of order $1/\gamma^2$.  The
expression (\ref{equiden}) is an important result since numerically
exact calculations of the path integral (\ref{equi1}) e.g.\ via Monte Carlo 
techniques 
become more expensive at lower temperatures. It further reveals that
for strong friction the equilibrium density consists of a
part which in phase space takes
the form of a classical distribution, however,  with an $\hbar$ dependent
$\langle p^2\rangle$  and a part with $\Lambda$
dependent quantum corrections (see also fig.~\ref{fig1}). 

{\it Quantum Fokker-Planck equation--}For the time evolution of the
density we basically apply the same kind 
of semiclassical analysis as briefly outlined above.
Since on the coarse grained time scale the dynamics
is effectively Markovian it suffices to calculate the propagating
function $J(\cdot)$ for a time step from $t$ to $t+\delta t$ where
$\delta t$ obeys $\gamma/\omega_0^2\gg\delta t\gg \hbar\beta,
1/\gamma, 1/\omega_c$. Our 
goal is to derive from this 
result a time evolution equation for the density which after a Wigner
transform gives rise to a quantum Fokker-Planck equation. 

We start by expanding the effective action up to second order in
$\bar{x}$ and $x(s)$. 
Specifically, for the potential terms in the real time actions one writes
 $V(r+x/2)-V(r-x/2)= V'(r) x+O(x^3)$.  
The complication that arises in evaluating 
the minimal action paths is that due to the correlations between
system and bath the corresponding equations of motions are
coupled. For anharmonic potential fields this necessitates in general
a numerical evaluation.
 However, here it turns out that one may write $\bar{q}_{\rm
ma}=r_i+\delta\bar{q}$ and $r_{\rm ma}=r_f+\delta r$ where
$\delta\bar{q}$ and  $\delta r$ are of order $\Lambda $ or
smaller. This way, $\delta\bar{q}$ is determined similar as for the
static case. By approximating $V'(r)= V'(r_f)+V''(r_f)\,
\delta r+O(\delta r^2)$ the real-time paths $\delta r$ and $x$ run
effectively in a harmonic oscillator potential with frequency
$\sqrt{V''(r_f)/M}$ subject to an external force $V'(r_f)$. The corresponding
part of the 
propagating function has been derived in \cite{grabert} to which we refer for
further details. Hence, to gain a time evolution equation for
$\rho(r_f,x_f,t)$ in the form $\dot{\rho}(t)={\cal L}\, \rho(t)$ we
look for an operator ${\cal L}={\cal L}(x_f,r_f,\partial/\partial
x_f,\partial/\partial r_f,t)$ being at most second order in the
coordinates and derivatives. Strong friction forces forward
and backward real time paths to run very close to eachother so that
quantum fluctuations---responsible for diffusion---reduce to Gaussian
noise.
 Below we will briefly discuss
higher order corrections. To proceed, we make a general  ansatz for
${\cal L}$ with coefficients specified by comparing $\dot{\rho}$ with
${\cal L}\rho$. Eventually, we switch to classical phase space
$\{x_f,r_f\}\to \{p,q\}$, i.e.\ $\rho(x_f,r_f,t)\to W(p,q,t)$, by the
replacement $x_f\to i\hbar 
\partial/\partial p$, $r_f\to q$ and $\partial/\partial x_f\to i
p/\hbar$, $\partial/\partial r_f\to \partial/\partial q$. This leads
 to the main result of this article, namely, the time evolution
equation for the phase space distribution $W(p,q,t)$ of a dissipative
quantum system in the strong friction limit
(\ref{eqcon})
\begin{eqnarray}
\frac{\partial}{\partial t}\, W(p,q,t)&=& \left\{
\frac{\partial}{\partial p}\, \left[V'_{\rm eff}(q)+\gamma\,
p\right]-\frac{p}{M}\frac{\partial}{\partial q}+\gamma\, \langle p^2\rangle\,
\frac{\partial^2}{\partial p^2}\right.\nonumber\\
&&\left. +\frac{\partial^2}{\partial q\partial p}\, \left[ 1/\beta+\Lambda
\, V''(q)-\langle p^2\rangle/M\right]\right\}\ W(p,q,t)\, .\label{timeevo}
\end{eqnarray}
Here, we have introduced an effective potential $V_{\rm
eff}=V+(\Lambda/2) \, V''$, and $\Lambda$ and also $\langle p^2\rangle=\Omega$
are specified in (\ref{coeff}). The first line on the r.h.s
coincides with a classical Fokker-Planck operator in an effective
force field \cite{risken}, the second line describes quantum
mechanical coupled $p$-$q$ diffusion. In the 
high temperature limit $\gamma\hbar\beta\to 0$ the quantum
Fokker-Planck equation (QFP) tends to the classical Kramers equation
\cite{risken}. For small but finite $\gamma\hbar\beta$ and in case
of a harmonic potential
the QFP coincides with the master equation gained by Haake and
Reibold \cite{haake}, but differs from the Caldeira-Leggett master
equation \cite{caldeira} by the $p$-$q$ diffusion
term \cite{deri}. However, while 
these known master equations are restricted (for $\gamma/\omega_0>1$) to
the range of small $\gamma\hbar\beta$, the new QFP is valid for {\em all}
$\gamma\hbar\beta$. We only mention here, that the above QFP is not of
Lindblad form due to the coarse graining procedure on which its
derivation is based.
 Of course, the equilibrium solution to (\ref{timeevo}) is given by the Wigner
transform of (\ref{equiden}). Let us 
briefly touch the question about higher order diffusion terms to
(\ref{timeevo}). For harmonic systems they do not occur so that
the QFP is in this sense exact \cite{talkner}. In case of anharmonic
potentials they result from non-Gaussian quantum
fluctuations attributed to higher than second order derivatives in
$V(q)$.  A rough 
estimate shows that
anharmonic terms in $x_f$ (leading to higher than second order
derivatives in $p$) in the crucial low temperature range
$\gamma\hbar\beta\gg 1$ are of order 
$1/[\hbar\gamma^{3/2} \log(\omega_c/\gamma)]$ compared to the leading terms.  

{\it Applications--}With the QFP at hand, we are now able to study
phase space properties 
of quantum systems in strongly condensed
phase. In particular, classically, due to the time scale separation the
strong friction limit allows for a reduction of the phase space
Kramers equation 
to the position space Smoluchowski equation \cite{risken,skinner}
which has been of great 
importance in a variety of systems in physics and chemistry (cf.\ the
Introduction).
Its generalization to the
low temperature quantum 
domain has been found recently in \cite{ankerhold} by focusing on the path
integral expression for the diagonal
part $\rho(q,q,t)$. Now, from the quantum
phase space dynamics this latter result can
 be rederived and especially the influence of inertia effects can
be explored. For this purpose we employ the projection operator techniques
invoked in \cite{skinner} to
systematically reduce the QFP to position space. Along these lines we
introduce the operators ${\cal P}=f_\beta(p) \int dp$ and ${\cal
Q}=1-{\cal P}$ where $f_\beta(p)$ is the normalized momentum
distribution in equilibrium according to (\ref{equiden}).  The next steps are
 straightforward and not presented here in detail. After some
algebra one arrives to order $\Lambda/\gamma^3$ at an equation for the
position distribution 
$n(q,t)=f_\beta^{-1} \, {\cal P}W$ of the form
\begin{equation}
\frac{\partial}{\partial t} n(q,t)= \frac{1}{\gamma M}
\frac{\partial}{\partial q}\, \left\{
1+\frac{1}{M \gamma^2}\left[V''_{\rm eff}+\Lambda
V'''\right]\right\}\, L_{\rm QSE}\, n(q,t)\, .\label{qse}
\end{equation}
$L_{\rm QSE}=V'_{\rm eff}+\partial/\partial q [1/\beta+\Lambda
V'']$ is the quantum Smoluchowksi operator already derived in
\cite{ankerhold} (slightly generalized to
all values of $\gamma\hbar\beta$). A classical ($\Lambda=0$)
inertia correction $\propto V''$ appears,  while
quantum fluctuations enter through third and forth order derivatives
of the potential. Overdamped quantum Brownian motion in position space
thus becomes much more sensitive to the details of the potential 
profile at lower temperatures. 

There is one non-trivial case where analytical results are available
and can be compared with 
exact ones, namely, the damped harmonic oscillator. As already
mentioned above, in the quantum case initial correlations between
system and reservoir are of crucial importance and render the
calculation even for the harmonic case quite 
cumbersome \cite{grabert}. This is particularly true for the strong
friction range 
where factorizing initial states cannot be used. We illustrate this
in detail by considering the relaxation of expectation values in
position $\langle q(t)\rangle$ and $\langle
q(t)^2\rangle$ (see fig.~\ref{fig2}).
Already for a moderate damping is the QFP solution for the mean position 
in very good agreement with 
the exact dynamics, while that based on a factorizing initial state is
completely off. Upon closer inspection one
finds that in the latter case the $q(0)$-dependence is suppressed by a
factor $1/\gamma^2$ compared to the former ones due to the lack of
 initial correlations. For the
mean square in position deviations between  the QFP result and the
exact one are almost invisible. Remarkably, the impact of quantum
fluctuations due  to $\Lambda$ and   
$\langle p^2\rangle$, in the QFP encoded in the $q$-$p$ diffusion
term, is quite substantial.

{\it Conclusion--}We have analyzed the time evolution of strongly
damped quantum systems in phase space starting from the exact path
integral expression in position space. 
A suppression of quantum fluctuations in position is associated with
enhanced fluctuations in momentum.
Our central result is a quantum
Fokker-Planck equation applicable from the classical high temperature
($\gamma\hbar\beta\ll 1 $) to the low temperature quantum
($\gamma\hbar\beta\gg 1$) range. This way, 
while in the weak friction limit master equations have been known for years,
here, we have given the missing complement for strong damping. 
 The QFP now opens the
door to explore phase space features of many systems 
well studied in classical physics also at lower temperatures,
e.g.\ Kramers rate theory, driven transport, or soft matter problems.

{\it Ackknowledgements--}
This work was supported  by the Nachkontaktprogramm of the Alexander
von Humboldt Foundation. Discussions with P. Pechukas are gratefully
acknowledged.

\begin{figure}
\caption{ $\Lambda$ as a measure for position fluctuations and $\Omega=\langle
p^2\rangle$ vs.\ inverse temperature for different
$\gamma$ and a Drude model with $\omega_c/\omega_0=50$. $\Lambda$ is
scaled by $\hbar/M\omega_0$, $\Omega$ by $\hbar 
M\omega_0$.} 
\label{fig1}
\end{figure}

\begin{figure}
\caption{ Relaxation of position expectation values for a damped harmonic
oscillator within a Drude model  $\omega_c/\omega_0=50$,
$\gamma/\omega_0=5$, $\omega_0\hbar\beta=1$. Initial values are for left panel:
$\langle p(0)\rangle/M \omega_0\langle q(0)\rangle=1$; for right panel:
$\langle q^2(0)\rangle=\langle p(0)^2\rangle/M^2\omega_0^2=\langle
q^2\rangle/4$ 
(other moments are zero) where $\langle
q^2\rangle $ is the quantum equilibrium variance.}
\label{fig2}
\end{figure}

\pagebreak

\setlength{\unitlength}{1cm}

\begin{picture}(0,17)

\put(-2.5,-6){
\makebox(22,22){
               \epsfysize=25cm
               \epsffile{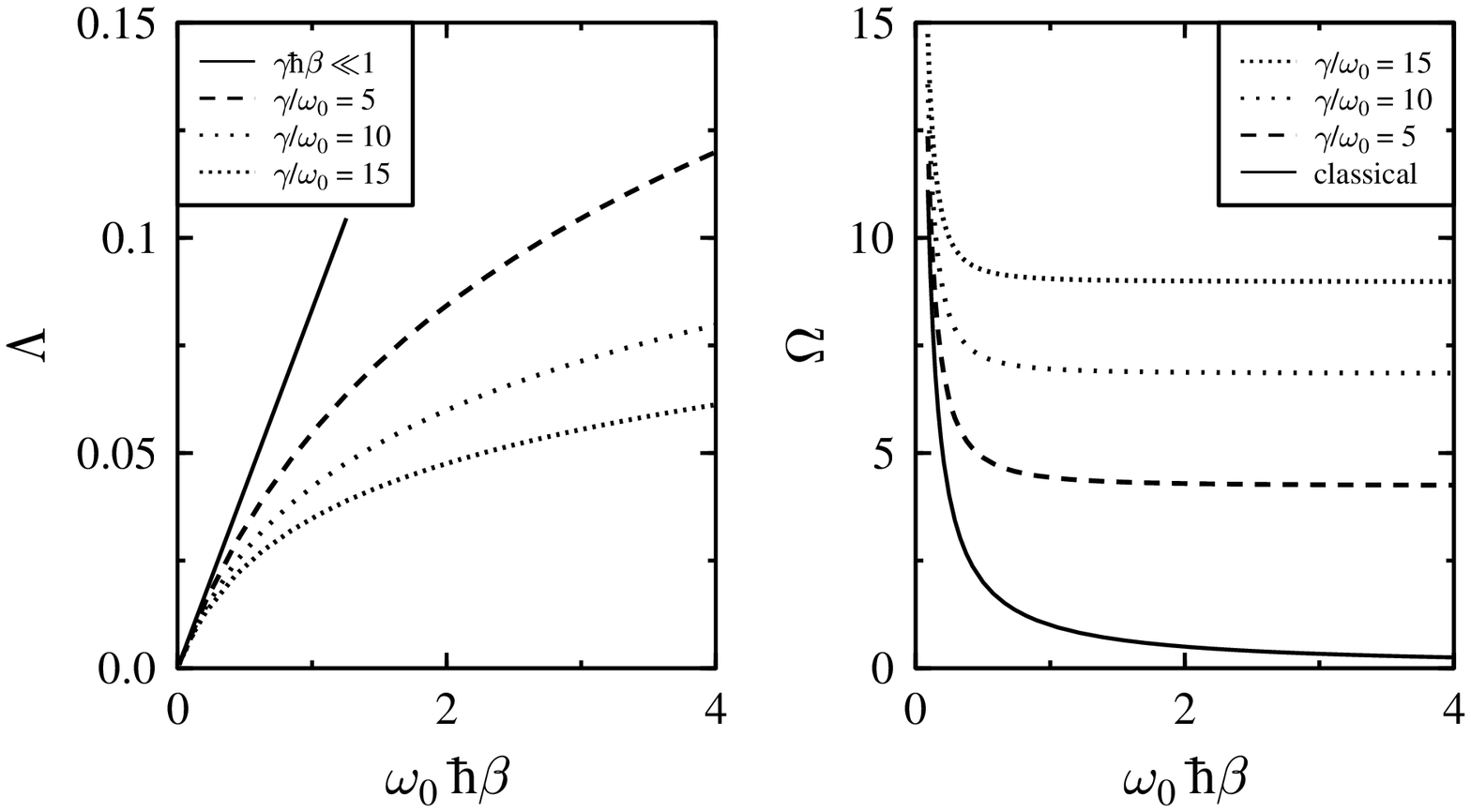
                        }
                          }
          }

\end{picture}

\pagebreak

\begin{picture}(0,17)

\put(-2.5,-6){
\makebox(22,22){
               \epsfysize=25cm
               \epsffile{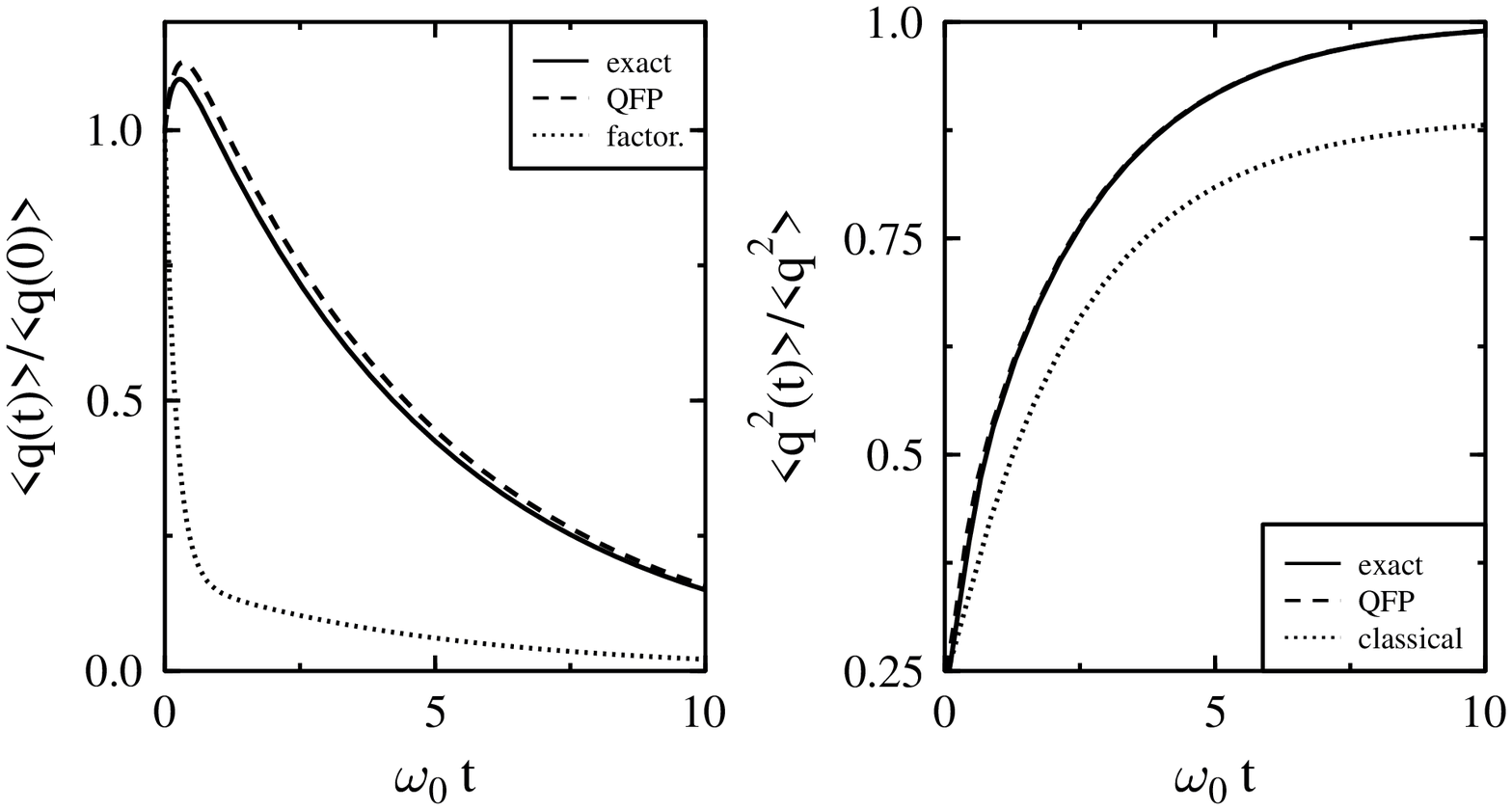
                        }
                          }
          }

\end{picture}


\begin{thebibliography}{99}
\bibitem{jung} L.\ Gammaitoni, P.\ H\"anggi, P.\ Jung, and F.\ Marchesoni,
Rev.\ Mod.\ Phys.\ {\bf 70}, 223 (1998).
\bibitem{fluc1} C.R.\ Doering, J.C.\ Gadoua, Phys.\ Rev.\ Lett.\ {\bf
69}, 251 (1990).
\bibitem{reimann} P. \ Reimann, Phys. Rep. {\bf 361}, 57 (2002). 
\bibitem{garg} A. Garg, J.N. Onuchic, and V. Ambegaokar,
J. Chem. Phys. {\bf 83}, 4491 (1985).
\bibitem{risken} H.\ Risken, {\it The Fokker Planck Equation},
(Springer, Berlin, 1984).
\bibitem{weiss} U.\ Weiss, {\it Quantum Dissipative Systems}, (Singapore,
World Scientific, 1999).
\bibitem{thorwart} M.\ Thorwart, M.\ Grifoni, and P.\ H\"anggi,
Phys. Rev. Lett. {\bf 85}, 860 (2000).
\bibitem{ankerhold} J.\ Ankerhold, P.\ Pechukas, and H.\ Grabert,
Phys. Rev. Lett. {\bf 87}, 086802 (2001).
\bibitem{anker} J. Ankerhold, Phys. Rev. E {\bf 64}, 060102 (R)
(2001).
\bibitem{baner} D. Banerjee, B.C. Bag, S.K. Banik, and D.S. Ray,
cond-mat/0205508 v1.
\bibitem{grabert} H.\ Grabert, P.\ Schramm, G.-L.\ Ingold, Phys.\ Rep.\
{\bf 168}, 115 (1988).
\bibitem{haake} F. Haake and R. Reibold, Phys. Rev. A {\bf 32}, 2462 (1985).
\bibitem{caldeira} A.O. Caldeira and A.J. Leggett, Physica A {\bf
121}, 587 (1983).
\bibitem{deri} Any consistent derivation of a master equation
from the path integral expression must include this term, see: R. Karrlein and
H. Grabert, 
Phys. Rev. E {\bf 55}, 153 (1997).
\bibitem{talkner} P. Talkner, Ann. Phys. (NY) {\bf 167}, 390 (1986).
\bibitem{skinner} J. L. Skinner and P.G. Wolynes, Physica A {\bf 96},
561 (1979).



\end{thebibliography}
\end{document}